\documentclass{PoS}
\usepackage{xspace}

\usepackage{cite}

\newcommand{\sNNtwohundred}{$\sqrt{s_{NN}}$ = $200$ GeV\xspace}
\newcommand{\sNNsixtytwo}{$\sqrt{s_{NN}}$ = $62$ GeV\xspace}
\newcommand{\sNNboth}{$\sqrt{s_{NN}}$ = $62$ GeV and $200$ GeV\xspace}

\newcommand{\stdassoc}{$1.5$ GeV/c $<$ $p_T^{associated}$ $<$ $p_T^{trigger}$\xspace}
\newcommand{\stdtrig}{$3.0$ $<$ $p_T^{trigger}$ $<$ $6.0$ GeV/c\xspace}
\newcommand{\pttrig}{$p_T^{trigger}$\xspace}
\newcommand{\ptassoc}{$p_T^{associated}$\xspace}
\newcommand{\npart}{$N_{part}$\xspace}

\newcommand{\pT}{$p_T$\xspace}

\newcommand{\zT}{$z_T$\xspace}

\newcommand{\jlc}{jet-like correlation\xspace}
\newcommand{\njet}{$Y_{Jet}$\xspace}

\newcommand{\ridge}{\textit{Ridge}\xspace}
\newcommand{\nridge}{$Y_{Ridge}$\xspace}

\newcommand{\pp}{$p+p$\xspace}

\newcommand{\Cu}{$Cu+Cu$\xspace}
\newcommand{\Au}{$Au+Au$\xspace}

\newcommand{\py}{PYTHIA\xspace}
\newcommand{\dAu}{$d+Au$\xspace}
\newcommand{\AplusA}{$A+A$\xspace}
\newcommand{\MeV}{MeV/c\xspace}
\newcommand{\GeV}{GeV/c\xspace}
\newcommand{\dphi}{$\Delta\phi$\xspace}
\newcommand{\deta}{$\Delta\eta$\xspace}
\newcommand{\vtwo}{$v_2$\xspace}
\newcommand{\nearside}{near-side\xspace}

\newcommand{\ns}{near-side\xspace}

\newcommand{\Fref}[1]{Fig.~\ref{#1}}
\newcommand{\Tref}[1]{Table~\ref{#1}}

\newcommand{\Cref}[1]{Chapter~\ref{#1}}

\newcommand{\roughly}{$\approx$\xspace}

\newcommand{\jl}{jet-like\xspace}

\newcommand{\jly}{jet-like yield\xspace}
\newcommand{\ry}{\ridge yield\xspace}

\newcommand{\pthat}{$\hat{p_T}$\xspace}

\newcommand{\etal}{{\it et al\/}\ }

\title{System size and energy dependence of the near-side of high-$p_T$ triggered di-hadron correlations in STAR}

\ShortTitle{System size and energy dependence of di-hadron correlations in STAR}

\author{\speaker{Christine Nattrass for the STAR collaboration}\\
        Yale University\\
        E-mail: \email{christine.nattrass@yale.edu}}

\abstract{Previous studies have indicated that the \ns peak of high-\pT triggered di-hadron correlations can be decomposed into two parts, a \jlc and the \ridge.  We present data from Cu+Cu and Au+Au collisions at \sNNboth, which should allow more robust tests of models.  The \jlc is narrow in both azimuth and pseudorapidity and has properties similar to those expected from vacuum fragmentation.  The yield of particles in the \jlc are presented for both systems and at both energies and compared to the yields expected from di-hadron correlations in \py 8.1.  The \ridge is narrow in azimuth but broad in pseudorapidity and roughly independent of pseudorapidity within STAR's acceptance.  Attempts have been made to explain the production of the \ridge component as coming from recombination, momentum kicks, Glasma flux tubes, and a plasma instability.  However, few models have attempted to quantitatively calculate the characteristics of the \ridge.  The yield in the \ridge is compared for all systems and energies.  The implications for measurements at the LHC are discussed in context of the data and models.}

\FullConference{High-pT Physics at LHC -09\\
		 February 4- 7 2009\\
		 Prague, Czech Republic}

\begin{document}
\section{Introduction}
Previous studies at RHIC demonstrated significant modification of the \ns peak of high-\pT triggered di-hadron correlations \cite{Joern,Jana}.  The \ns can be decomposed into two components, a \jlc and the \ridge.  The \jlc is similar to high-\pT triggered di-hadron correlations observed in \pp, \dAu, and \py.  It is narrow in both azimuth and pseudorapidity and its particle composition is similar to the inclusive particle ratios in \pp at the same collision energy and particle momenta \cite{Jana,Cristina}.  

The \ridge is a novel feature not observed in \pp and \dAu collisions and not explained by simple models such as \py.    STAR has done extensive studies of the properties of the \ridge; other relevant measurements are discussed in \cite{JanaPrague,OlgaPrague}.    Although this paper focuses on high-\pT triggered correlations, a "soft" \ridge has been observed by the STAR collaboration in untriggered correlations as well \cite{MiniJets,MikeD}.  Many mechanisms have been proposed for the production of the \ridge but there have been few quantitative comparisons with the data to date.

The momentum kick model assumes an initial distribution of medium partons which is roughly independent of pseudorapidity.  Collisions with a hard parton create a correlation in azimuth between the hard parton and the medium partons \cite{MomentumKick}.   Several mechanisms for the production of the \ridge involve hydrodynamics.  One model explains the \ridge as resulting from a parton wind due to longitudinal flow \cite{LongFlow}.  It is also proposed that the \ridge arises due to a combination of a bias for the trigger to be near the surface of the medium and radial flow \cite{Sergei,SergeiLatest,Shuryak}.  Recent calculations in a full hydrodynamical model also show a \ridge \cite{Jun}.  Several models explain the \ridge as a correlation in pseudorapidity created by plasma instabilities from fluctuations in strong fields early in the collision \cite{PlasmaInstability,PlasmaInstability1,PlasmaInstability2,PlasmaInstability3,McLerran1,McLerran2}.  Some of these models are compared to the untriggered data, however, these models have been hypothesized to be able to explain the high-\pT triggered \ridge \cite{McLerran1,McLerran2}; these models also need radial flow to produce a \ridge large enough to be comparable to the data.
\section{Analysis method}
Data from the STAR detector from \Au collisions at \sNNsixtytwo are from RHIC's fourth year run (2004) and data from \Cu collisions at \sNNboth are from RHIC's fifth year run (2005.)  Details of the STAR detector can be found in \cite{STARNIM}.  These data are compared to previous studies done in the \Au data at \sNNtwohundred from RHIC's fourth year run \cite{Jana}.

In a given event, a high-\pT track, called the trigger particle, is selected and the distribution of lower momentum particles, called associated particles, relative to that track in \deta and \dphi is determined.  Multiple tracks in a single event may be counted as trigger and associated particles.  Since these measurements are limited by statistics and the data are symmetric about \dphi = 0 and \deta = 0, the data are reflected about \dphi = 0 and \deta = 0 to minimize statistical fluctuations.  The single track efficiency is dependent on particle type, collision system and energy, \pT, and centrality and the correction for associated particles is applied on a track-by-track basis.  The final distribution is normalized by the number of trigger particles, so no correction for the efficiency of reconstructing the trigger particle is necessary.  Detector acceptance is corrected for by selecting a random trigger from the distribution of trigger particles in ($\phi$,$\eta$) and mixing it with a random associated particle from the distribution of associated particles in ($\phi$,$\eta$).  This is done for each bin in \pT and centrality.  This is used to calculated the correction for detector acceptance as a function of (\dphi,\deta).  This is done for the full sample of events to give the average distribution of associated particles relative to a high-\pT trigger particle $\frac{1}{N_{trigger}}\frac{d^2N}{d\Delta\phi d\Delta\eta}$.

The \ridge was previously observed to be approximately independent of \deta within the acceptance of the STAR TPC while the \jlc is confined to $|$\deta$|$ $<$ 0.75 \cite{Joern,MeSQM}.  This is used to separate the \jlc and the \ridge.  The yield, number of particles associated with each trigger particle within limits on \ptassoc and \pttrig, is studied.

\subsection{Determination of \jly}

To determine the \jly, \njet, the projection of the distribution of particles $\frac{d^2N}{d\Delta\phi d\Delta\eta}$ is taken in two different ranges in pseudorapidity:
\begin{eqnarray}
  \frac{dY_{ridge}}{d\Delta\phi} & = & \frac{1}{N_{trigger}} \int\limits_{-1.75}^{-0.75} \frac{d^2N}{d\Delta\phi d\Delta\eta} d\Delta\eta 
   + \frac{1}{N_{trigger}} \int\limits_{0.75}^{1.75} \frac{d^2N}{d\Delta\phi d\Delta\eta} d\Delta\eta\\ \label{ridgeeqtn1}
\frac{dY_{jet+ridge}}{d\Delta\phi} &= &\frac{1}{N_{trigger}} \int\limits_{-0.75}^{0.75} \frac{d^2N}{d\Delta\phi d\Delta\eta} d\Delta\eta\label{jeteqtn1}
\end{eqnarray}
where the former contains only the \ridge and the latter contains both the \jlc and the \ridge.  The \jl yield on the \nearside is the integral over $-1<$ \dphi $<1$:
\begin{equation}
Y_{jet}^{\Delta\phi} = \int\limits_{-1}^{1} ( \frac{dY_{jet+ridge}}{d\Delta\phi} - \frac{0.75}{1}  \frac{dY_{ridge}}{d\Delta\phi} ) d\Delta\phi. \label{jetphieqtn}
\end{equation}
The factor in front of the second term is the ratio of the \deta width in the region containing the \jlc and the \ridge to the width of the region containing only the \ridge.  The \njet is determined by fitting a Gaussian to the peak at \dphi \roughly 0 after subtracting the \ridge.  Since the data have been reflected about \dphi = 0 and \deta = 0, the projections in \dphi and \deta and the bin counting are done over half of the range and then scaled up to ensure correct error propagation.

There is a slight loss of particles at small (\dphi,\deta) due to track merging.  The correction for this effect is in progress.  It affects only the \jlc and leads to a loss of \roughly 10\% of the \jly at the lowest \pttrig in central \Au, where it is largest \cite{Marek}.

\subsection{Determination of \ridge yield}
There is a large combinatorial background which is dependent on \dphi given by
\begin{equation}
\frac{dY_{bkgd}}{d\phi} = B (1 + 2 \langle v_2^{trigger}\rangle \langle v_2^{associated}\rangle \cos(2\Delta\phi) )\label{bkgdeqtn}
\end{equation}
where \vtwo is the second order harmonic in a Fourier expansion of the momentum anisotropy relative to the reaction plane \cite{Janavtwo}.  Systematic errors come from the errors on B, $\langle v_2^{trigger}\rangle$ and $\langle v_2^{associated}\rangle$.  \vtwo is determined from independent measurements.  It is assumed that \vtwo is the same at a given multiplicity for events with a trigger particle as for those without a trigger particle and that \vtwo is independent of $\eta$, a reasonable assumption within the acceptance of the STAR TPC based on the available data \cite{phobosFlow1,phobosFlow2}.  For each data set $v_2(p_T)$ was fit to the measured data for each centrality bin to determine $\langle v_2^{trigger}\rangle$ and $\langle v_2^{associated}\rangle$.  Details of the \vtwo subtraction are given in \cite{Jana} for \Au collisions at \sNNtwohundred and in \cite{MeSQM} for \Cu collisions at \sNNtwohundred.  For \Cu collisions at \sNNsixtytwo, \vtwo using the reaction plane as determined from tracks in the Forward Time Projection Chamber was used as the nominal value and the lower bound was determined from a multiplicity-dependent approximation as described for \sNNtwohundred in \cite{MeSQM}.  \vtwo and its systematic errors were taken from \cite{AuAuSixtyTwoFlow} for \Au collisions at \sNNsixtytwo.  B is determined using the Zero Yield At Minimum (ZYAM) method \cite{starZYAM}.

To determine the yield of the \ridge, \nridge, the di-hadron correlation $\frac{d^2N}{d\Delta\phi d\Delta\eta}$ is projected over the entire \deta region.  To minimize the effects of statistical fluctuations in the determination of the background, $\frac{dY_{bkgd}}{d\phi}$, \njet is subtracted after the projection in \deta:
\begin{equation}
Y_{Ridge} = 1/N_{trigger} \int\limits_{-1.75}^{1.75} \int\limits_{-1}^{1} (\frac{d^2N}{d\Delta\phi d\Delta\eta} - \frac{dY_{bkgd}}{d\phi}) d\Delta\phi d\Delta\eta - Y_{Jet}^{\Delta\eta}. 
\end{equation}
The integration over \dphi is done fitting a Gaussian to the peak around \dphi \roughly 0.

\section{Results}

\subsection{The \jlc}

\begin{figure}
\begin{center}
\resizebox{8.8cm}{!}{%
  \includegraphics{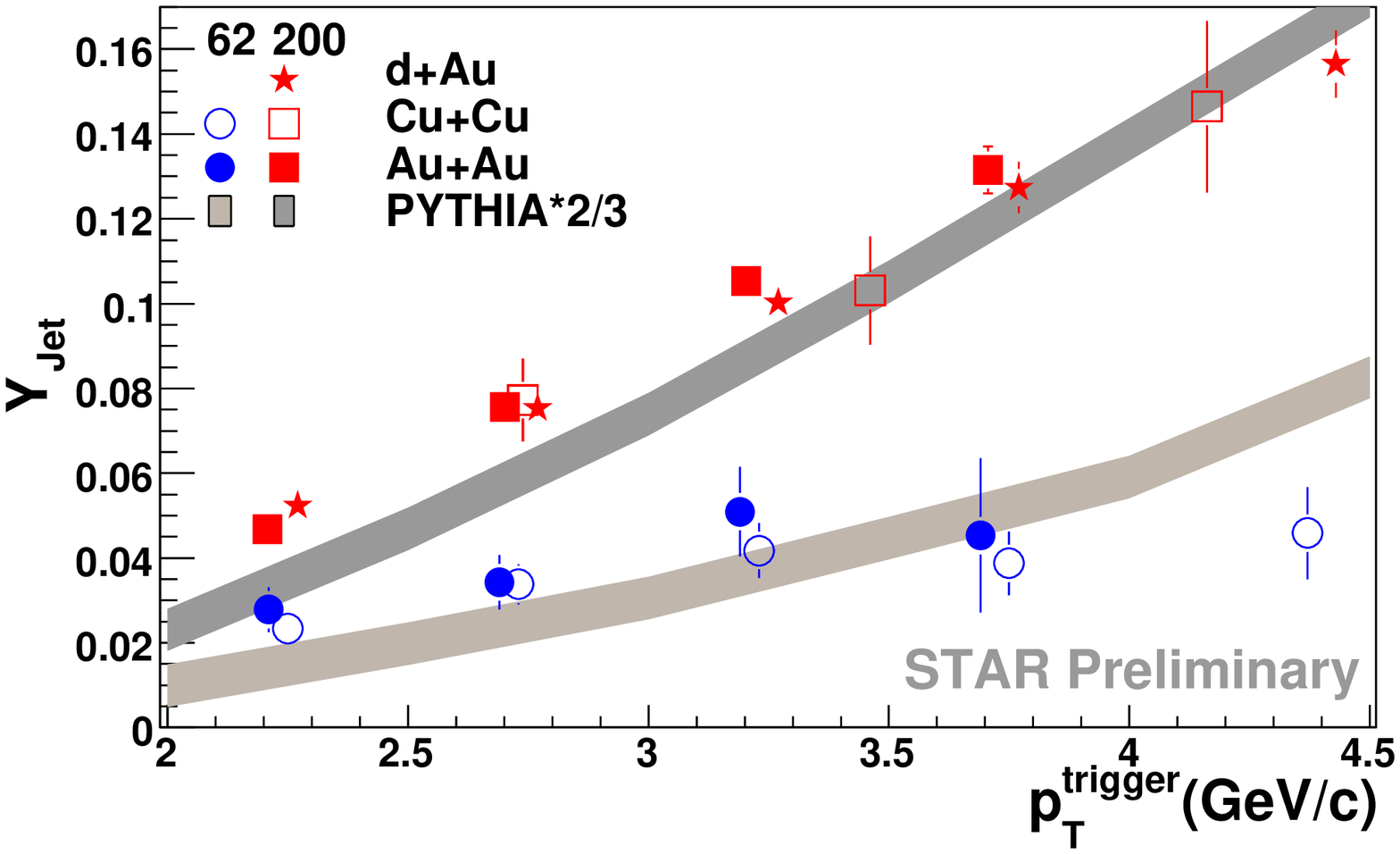}
}
\caption{
\pttrig dependence of the \njet for \Cu and \Au at \sNNsixtytwo and \dAu, \Cu, and \Au at \sNNtwohundred for \stdassoc compared to the yield from PYTHIA scaled by 2/3.
}
\label{TrigPt}
\end{center}
\end{figure}

\Fref{TrigPt} shows the \njet as a function of \pttrig for \Cu and \Au at \sNNsixtytwo and \dAu, \Cu, and \Au at \sNNtwohundred.  For a given collision energy, there is no difference within statistical errors between the different collision systems.  This would be expected if the \jlc is dominantly produced from vacuum fragmentation.  These data are compared to the charged particle yields expected from \py 8.1 using the default settings \cite{PYTHIAManual}, scaled by 2/3.  Although the yield is off by a factor of 2/3, the shape of the \pttrig dependence is described well by \py.

\begin{figure}
\begin{center}
\resizebox{12cm}{!}{%
  \includegraphics{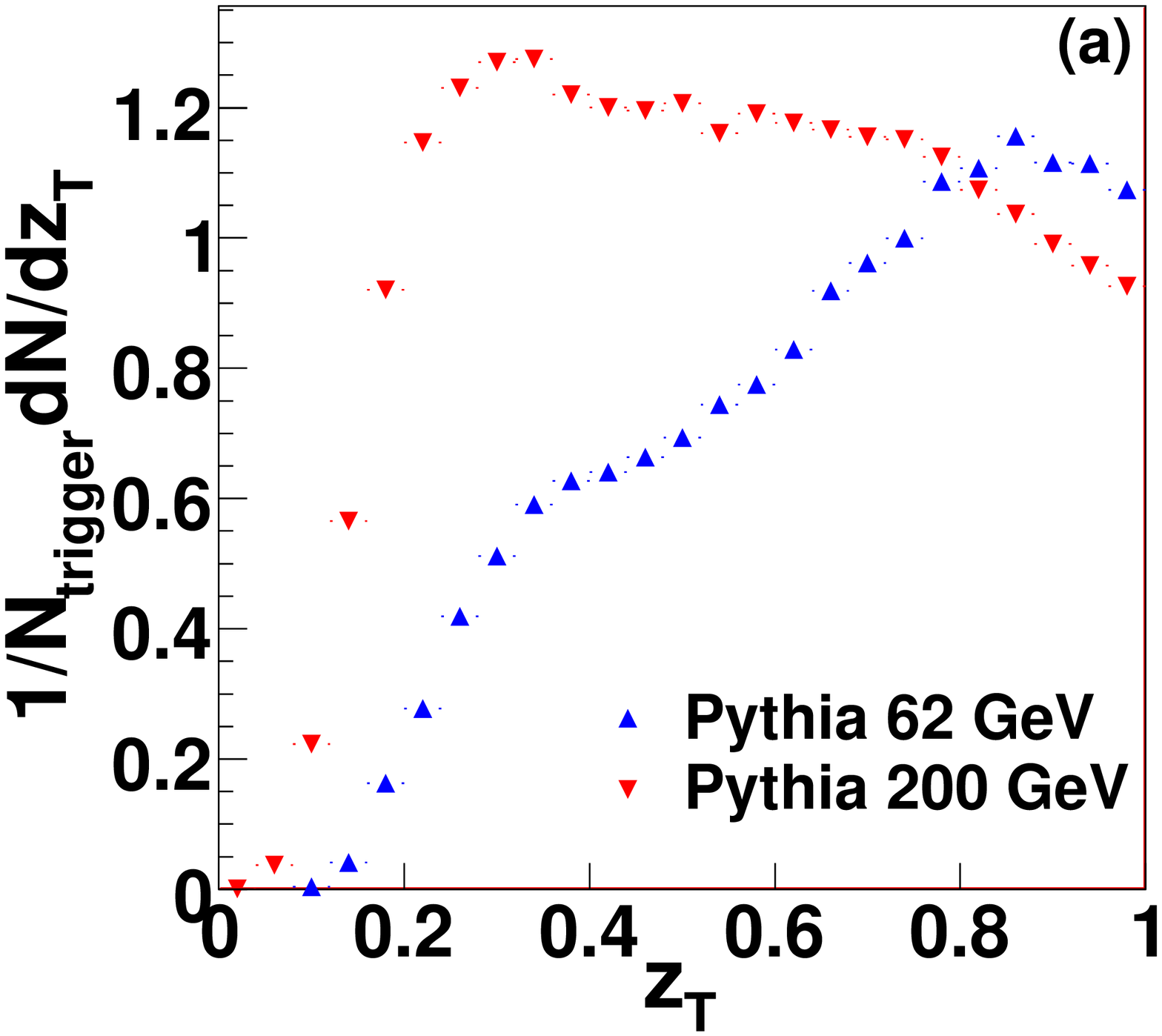}
  \includegraphics{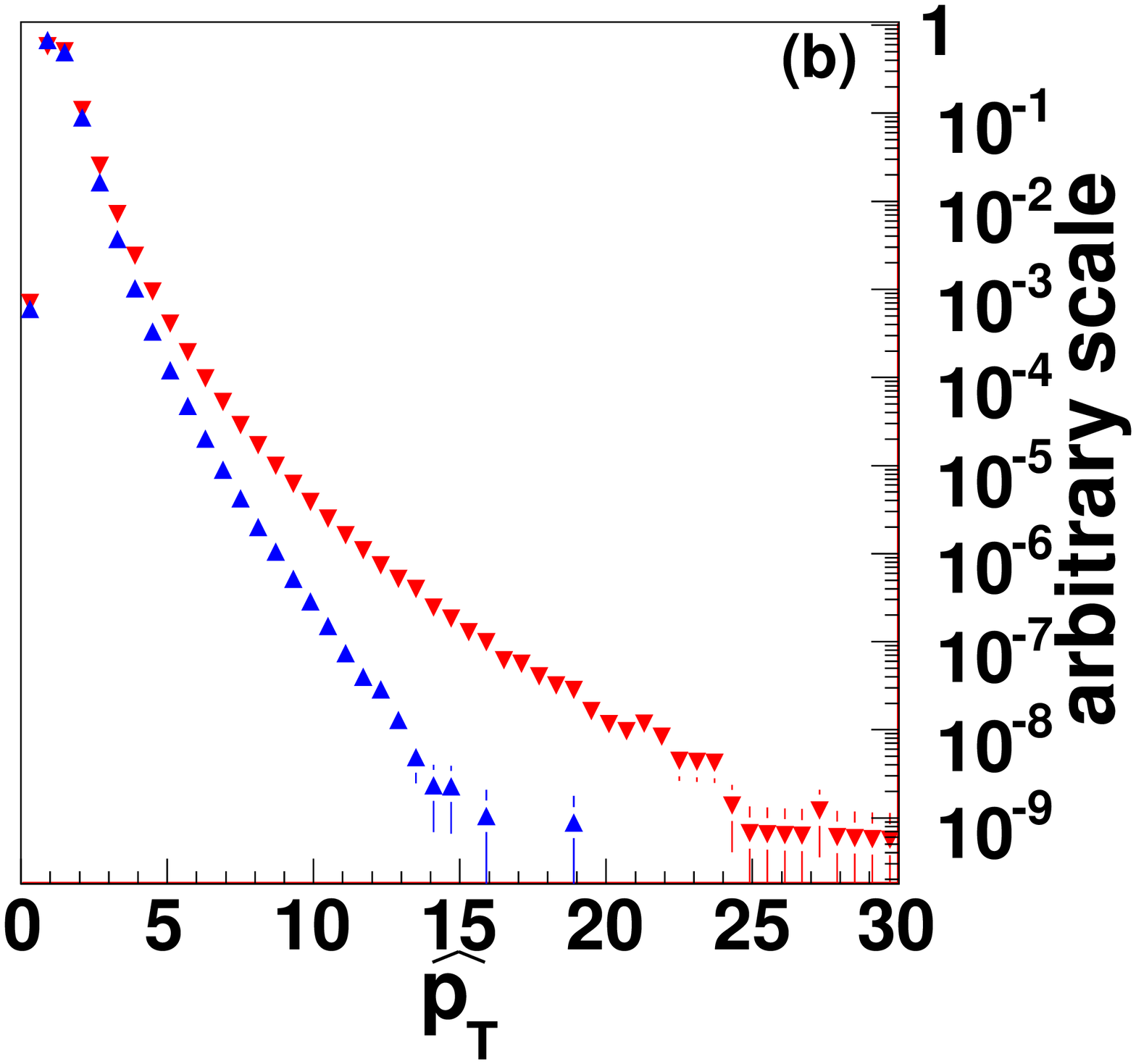}
}
\caption{(a) distribution of trigger particles \zT and (b) \pthat distribution in PYTHIA at \sNNsixtytwo from PYTHIA 8.1 for \stdassoc and \stdtrig at \sNNsixtytwo and \sNNtwohundred.}
\label{PYTHIA}       
\end{center}
\end{figure}

The difference between \sNNboth is also described well by \py. \Fref{PYTHIA}(a) shows the \zT = \pT/\pthat distribution in \py and \Fref{PYTHIA}(b) shows the distribution of \pthat, the momentum transfer of the hard scattering, at both energies.  The \pthat is the parameter in PYTHIA for the transverse momentum in the hard subprocess \cite{PYTHIAManual}.  For the same \pttrig, a higher average \zT is selected in collisions at \sNNsixtytwo, meaning that the trigger particle takes a greater percentage of the energy.  The slope of \pthat is considerably steeper in \sNNsixtytwo than in \sNNtwohundred so for a fixed \pttrig the trigger particle is more likely to have come from a lower energy parton. The large difference between \njet at \sNNboth can then be understood as arising from the difference in the underlying jet spectra which lead to the di-hadron correlations.

\begin{figure}
\begin{center}
\resizebox{8.8cm}{!}{  \includegraphics{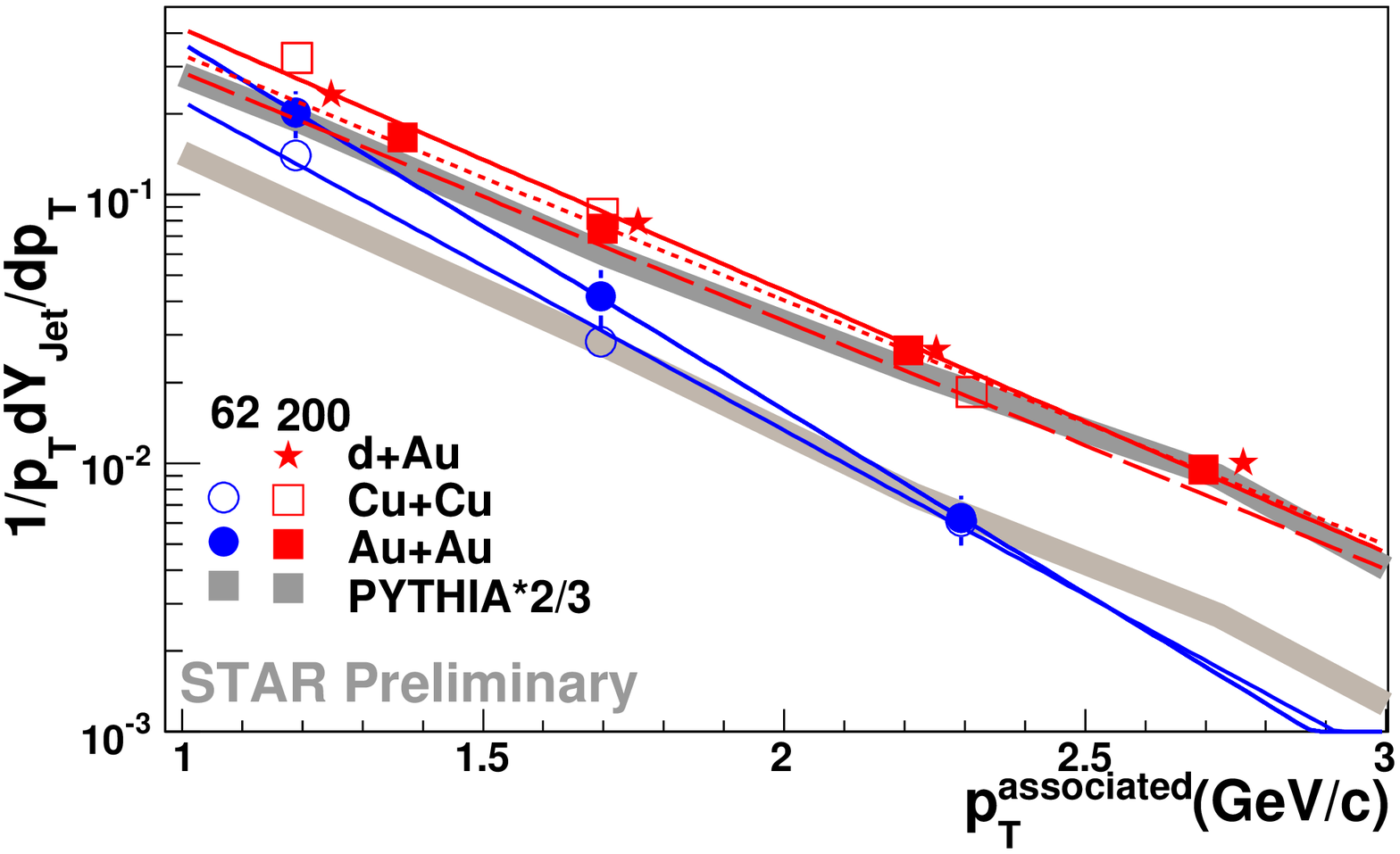} }
\caption{\ptassoc dependence of \njet for \Cu and \Au at \sNNsixtytwo and \dAu, \Cu, and \Au at \sNNtwohundred for \stdtrig compared to the yield from PYTHIA scaled by 2/3.  The inverse slope parameters from fits of an exponential to data and to PYTHIA are given in Table 1.}
\label{AssocPt}
\end{center}
\end{figure}

The \ptassoc dependence of \njet is shown in \Fref{AssocPt} for \Cu and \Au at \sNNsixtytwo and \dAu, \Cu, and \Au at \sNNtwohundred compared to the yield from PYTHIA scaled by 2/3.  The inverse slope parameters from fits of the data are given in \Tref{mytable}.  The data for \Cu and \Au collisions are within errors at a given collision energy.  \py overestimates the inverse slope parameter of the spectra of particles in the \jlc for collisions at \sNNsixtytwo.

\begin{table}
\begin{center}
\caption{Inverse slope parameter k (\MeV) of \ptassoc for fits of data in Table 1 to $A e^{-p_{T}/k}$.  The inverse slope parameter for inclusive $\pi^-$ are from a fit to the data in \cite{Pion} as a function of \pT above 1.0 \GeV.  Statistical errors only.}
\begin{tabular}[b]{c|c|c}
\hline
 & \sNNsixtytwo & \sNNtwohundred\\
   & h-h & h-h\\ \hline
Au \ridge&  & 438 $\pm$ 4   \\ 
\Au \jlc & 317 $\pm$ 26 & 478 $\pm$ 8 \\
\Cu \jlc & 355 $\pm$ 21 & 445 $\pm$ 20 \\ 
\dAu \jlc &  & 469 $\pm$ 8 \\ \hline
PYTHIA & 424 $\pm$ 5 & 473 $\pm$ 3\\ \hline
Inclusive $\pi^-$ & 280.9 $\pm$ 0.4 & 330.9 $\pm$ 0.3\\ \hline
\end{tabular}\label{mytable}
\end{center}
\end{table}

\begin{figure}
\begin{center}
\resizebox{8.8cm}{!}{%
  \includegraphics{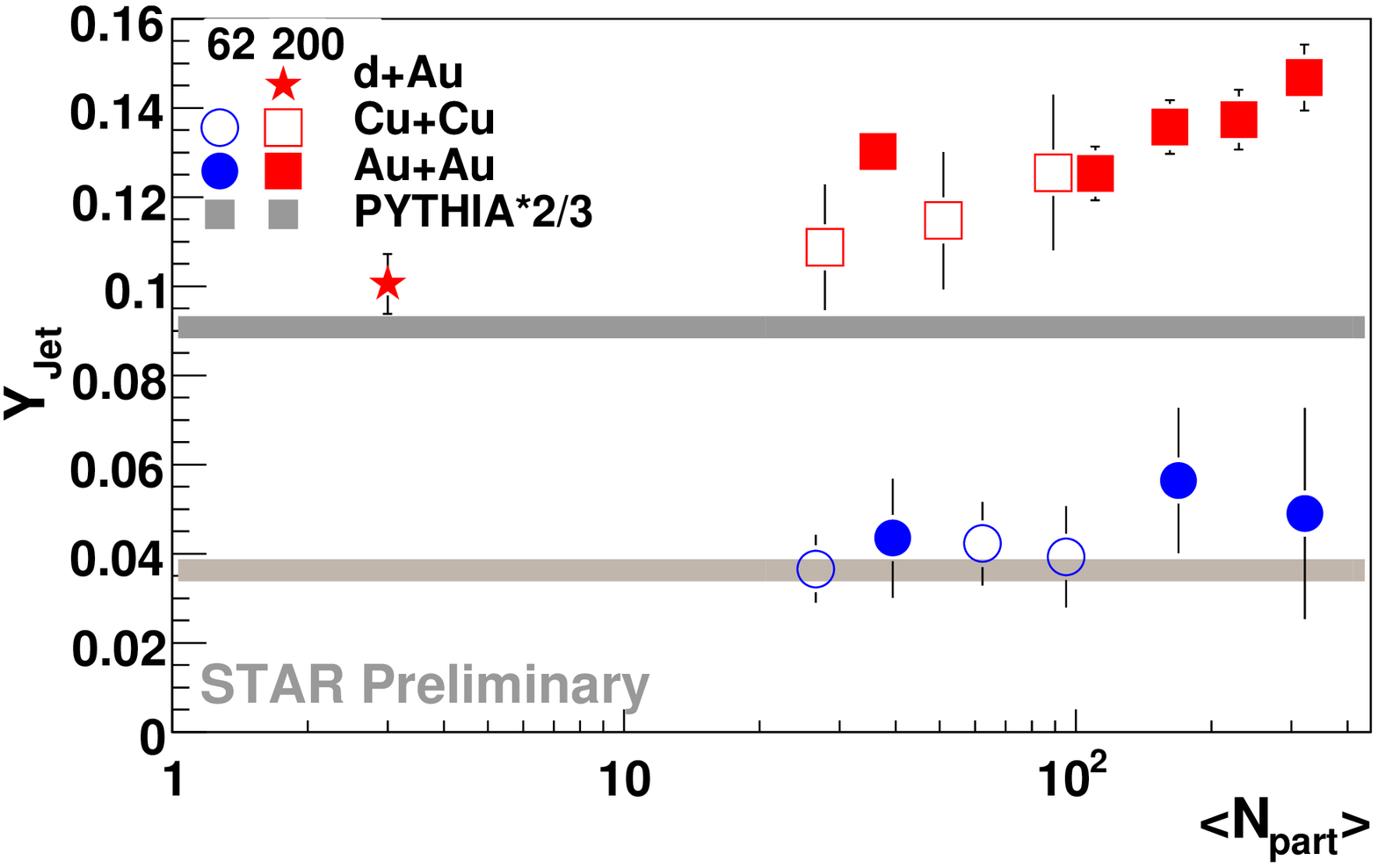}
}
\caption{\npart dependence of the \njet for \Cu and \Au at \sNNsixtytwo and \dAu, \Cu, and \Au at \sNNtwohundred for \stdtrig and \stdassoc compared to the yield from PYTHIA.}
\label{JetNpart}       
\end{center}
\end{figure}

\njet as a function of \npart is shown in \Fref{JetNpart}.  At a given \npart, the data in \Cu and \Au collisions are within errors at both \sNNsixtytwo and \sNNtwohundred.  There is a slight increase with \npart from \dAu to central \Au.  This increase may either be a slight modification of the \jlc or be a misidentification of part of the \ridge into the \jlc because of the inherent assumption in the method for determining the yield that the \ridge is independent of pseudorapidity.

\subsection{The \ridge}

\begin{figure}
\begin{center}
\resizebox{8.8cm}{!}{%
  \includegraphics{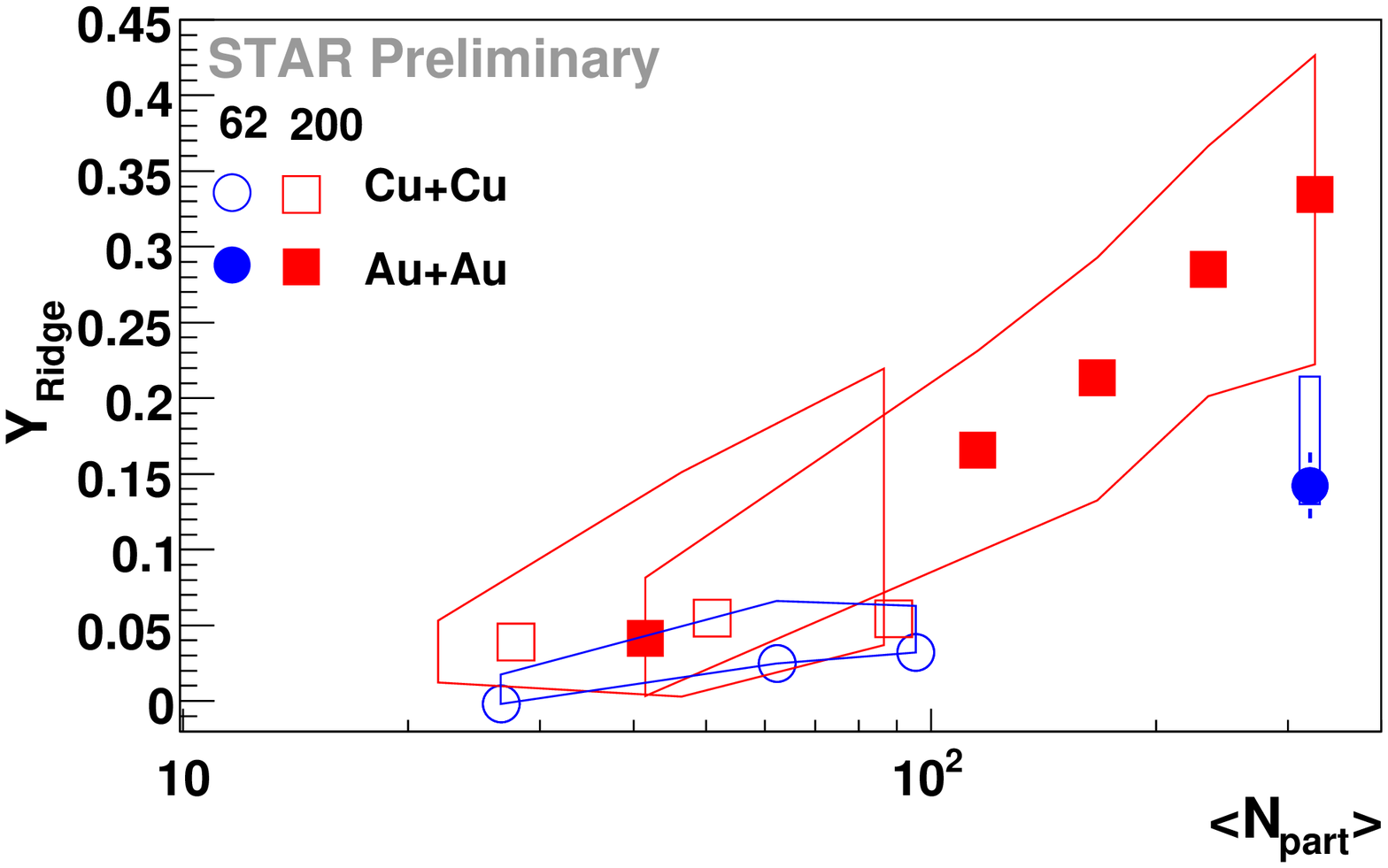}
}
\caption{\nridge dependence on \npart for \sNNsixtytwo and\sNNtwohundred for \stdtrig and \stdassoc.}
\label{RidgeNpart}       
\end{center}
\end{figure}

\begin{figure}
\begin{center}
\resizebox{8.8cm}{!}{%
  \includegraphics{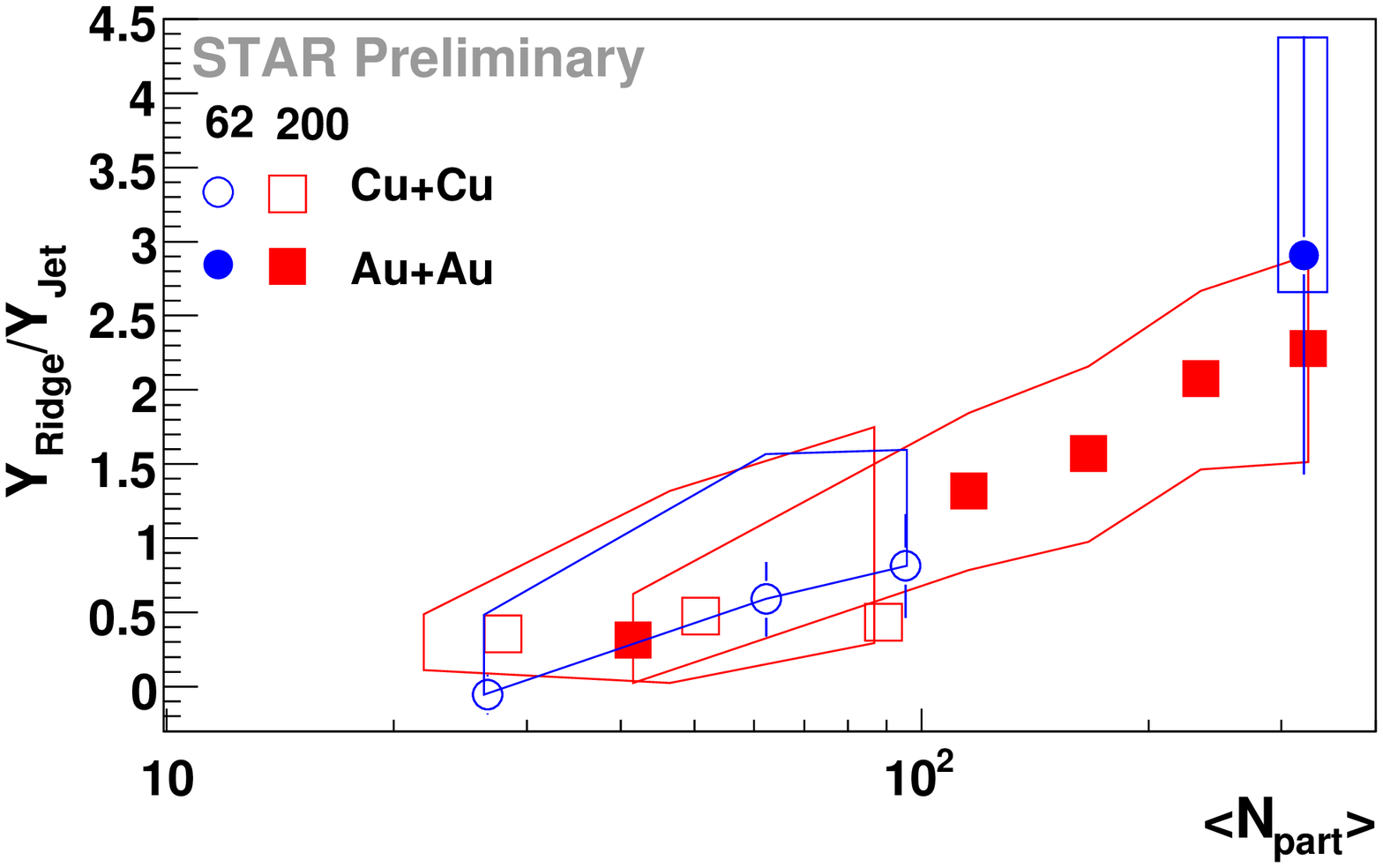}
}
\caption{\nridge/\njet dependence on \npart for \sNNsixtytwo and\sNNtwohundred for \stdtrig and \stdassoc.}
\label{RidgeRatio}       
\end{center}
\end{figure}

The \ry as a function of \npart is shown in \Fref{RidgeNpart}.  At a given \npart, the data from \Cu and \Au collisions at the same collision energy are within errors.  The \ry is considerably smaller in collisions at \sNNsixtytwo than those at \sNNtwohundred, just like the \jly.  An interesting trend is observed for the \ry; \Fref{RidgeRatio} shows that the ratio of the \ry to the \jly as a function of \npart for both energies is the same.

\section{Discussion and implications for the LHC}
The data imply that the \jlc is produced dominantly by fragmentation.  No collision system dependence is observed.  \njet increases slightly as a function of \npart; this may come from slight modification of the \jlc in \AplusA collisions or it may be an artifact of the method for separating the \jlc and the \ridge.  \py describes the shape of the \pttrig and collision energy dependence of the \jlc well, implying that the shape of the \pttrig and \ptassoc dependence is dominated by the kinematic cuts on \ptassoc and \pttrig.  This means that simple models such as \py can be used to understand the effects of the kinematic cuts on the distribution of jet energies which lead to the \jlc, providing insight into measurement of the \ridge.

Most models for the \ridge would qualitatively predict a \ridge at the LHC, however, few quantitative descriptions of the energy dependence of the \ridge have been made.  The Momentum Kick model can describe the energy dependence of the data at RHIC \cite{WongEnergy}.  Plasma instability models would also predict a \ridge, however, few quantitative comparisons of the models to the data have been done and there have been no attempts at describing the collision energy dependence \cite{PlasmaInstability,PlasmaInstability1,PlasmaInstability2,PlasmaInstability3,McLerran1,McLerran2}.  Models dependent on hydrodynamics \cite{LongFlow,Sergei,Jun} would also predict a \ridge at the LHC, however, no quantitative predictions have been made.  If collective flow is larger at the LHC, the \ridge would also be expected to be bigger.

A straightforward extrapolation from the data at RHIC can provide estimates for the expectations for measurements of the \ridge at the LHC.  \Fref{RidgeRatio} showed that, at least for the kinematic cuts used in these studies, the \nridge/\njet ratio is the same for collisions at \sNNsixtytwo and \sNNtwohundred.  If this is also true at LHC energies, the \nridge/\njet ratio will be the same for the same kinematic cuts.  The combinatorial background will be much larger at 5.5 TeV for \stdassoc and \stdtrig the dominant systematic error at RHIC is due to \vtwo subtraction, so measurements of the \ridge for the same kinematic cuts may be difficult if \vtwo at the LHC is comparable to that at RHIC.  The combinatorial background can be reduced by increasing \pttrig and \ptassoc.  \Fref{TrigPt} shows that \njet increases rapidly with \pttrig.  \nridge is roughly independent of \pttrig \cite{Joern}.  This means that increasing \pttrig decreases \nridge/\njet.  It was shown in \cite{Joern} that the spectrum of particles in the \ridge is softer than those in \njet.  This means that increasing \pttrig will also decrease \nridge/\njet.  Measurements of the \ridge at the LHC may be difficult because of the large combinatorial background if these naive extrapolations of RHIC data are correct.

\section{Conclusions}
Measurements at RHIC imply that the \jlc is dominantly produced by vacuum fragmentation.  There have been numerous measurements of the \ridge and many production mechanisms have been proposed, however, there have been few quantitative comparisons with the data.  Both the data and models imply that the \ridge should be present at the LHC.  With additional theoretical insight on the collision energy dependence of the \ridge, these measurements should be able to provide a better understanding of the origin of the \ridge.


%

\end{document}